# The AI Regulatory Readiness Index ARRI: Assessing Cross-Jurisdictional Legal Preparedness for AI in Telecommunications

Avinash Agarwal[a,*], Peeyush Agarwal[b], Manisha J. Nene[c]

[a]Department of Telecommunications, Ministry of Communications, Sanchar Bhawan, New Delhi, 110001, Delhi, India
[b]Netaji Subhas University of Technology, Dwarka, New Delhi, 110078, Delhi, India
[c]Defence Institute of Advanced Technology, Ministry of Defence, Girinagar, Pune, 411025, Maharashtra, India

**Abstract**

As Artificial Intelligence (AI) becomes increasingly embedded in critical telecommunications infrastructure, existing legal frameworks remain ill-equipped to address the distinct risks this development introduces. This paper proposes the **AI Regulatory Readiness Index (ARRI)**, a reproducible instrument for doctrinally assessing the legal preparedness of national frameworks to govern AI in critical digital infrastructure, and applies it across ten jurisdictions spanning five continents. ARRI comprises seven indicators across three dimensions: substantive AI-specific obligations, operational safeguards, and governance coordination, scored on a four-point ordinal scale and aggregated to a normalised 0–100 index. Legal instruments in force as of 28 February 2026 are assessed across telecommunications, cybersecurity, data protection, and AI governance domains. The study finds that global AI regulatory readiness in telecommunications remains concentrated in the lower range, with a mean ARRI score of 34 and a median of 26.5. AI incident reporting and risk classification emerge as the most acute and near-universal gaps, with binding legal definitions of AI-specific incidents largely absent across the legal frameworks applicable to telecommunications in the jurisdictions studied. ARRI scores diverge systematically from existing composite indices. For example, Indonesia achieves ITU Global Cybersecurity Index Tier 1 status yet scores 19 under ARRI, demonstrating that cybersecurity readiness and AI regulatory readiness are legally distinct conditions that existing frameworks conflate. The ten jurisdictions are classified into **five regulatory archetypes**, and a **normative minimum standards framework** is proposed, anchoring baseline AI governance readiness at an ARRI score of 67. ARRI is designed to be sector-portable and applicable beyond telecommunications to energy, healthcare, and transport infrastructure.

## 1. Introduction

The rapid integration of Artificial Intelligence (AI) into the foundational systems of modern society marks a paradigm shift, particularly within critical infrastructure. Critical infrastructures are defined as those indispensable systems and processes vital for societal functioning, with individual governments typically determining their designation [1]. While lauded for enhancing efficiency, this data-driven transition introduces inherent risks, including sophisticated adversarial attacks, biases, and system failures with potentially catastrophic consequences [2]. The inherent characteristics of AI, such as autonomous learning and unpredictability, introduce a new suite of challenges that can undermine the resilience and integrity of these vital systems.

The telecommunications sector stands out as a crucial component, forming the backbone of modern digital ecosystems. It leverages AI across a wide range of applications, including real-time cybersecurity through threat intelligence and risk assessment, predictive maintenance, network performance optimization, and network traffic analysis [3, 4]. However, this deep integration also introduces unique vulnerabilities. While traditional cybersecurity frameworks, such as those from the International Telecommunication Union (ITU), primarily address conventional threats [5, 6, 7], they may overlook a new class of non-malicious, systemic risks inherent to AI systems. Recent scholarship has highlighted this conceptual gap, arguing for a distinct legal categorization of *'Telecommunication AI Incidents'* to capture these novel failures, which include gradual model drift leading to network degradation, algorithmic bias in spectrum allocation causing discriminatory service quality, and emergent, unpredictable behaviour in self-organising networks [8]. Consequently, most existing telecommunications regulations remain silent on these purely AI-driven risks. Noteworthy that many existing legal frame-

*Corresponding author. The views expressed here are personal and do not necessarily reflect those of any organisation or institution.

*Email addresses:* avinash.70@gov.in (Avinash Agarwal), peeyush2478@gmail.com (Peeyush Agarwal), mjnene@diat.ac.in (Manisha J. Nene)

works, particularly national telecom acts, were drafted before the widespread integration of AI and thus often concentrate on traditional concerns like unauthorized access and data breaches. Furthermore, current regulatory paradigms tend to prioritize immediate incident response over systematic data collection, which is essential for model improvement and proactive risk mitigation.

While this substantive gap is clear, a more profound methodological challenge persists in legal scholarship: the absence of a structured, reproducible framework for assessing and comparing the AI regulatory readiness of different jurisdictions. Existing composite indices, such as the ITU Global Cybersecurity Index (GCI) [9] and the Oxford Insights Government AI Readiness Index (GARI) [10], provide valuable cross-national comparisons but do not disaggregate legal preparedness within specific critical infrastructure sectors or systematically distinguish between binding legal obligations and soft-law instruments. This paper addresses that gap by proposing the **AI Regulatory Readiness Index (ARRI)**, a sector-specific, doctrinally-grounded assessment instrument designed to evaluate the legal preparedness of national frameworks.

By applying the ARRI to ten jurisdictions, this paper moves beyond description to provide a structured, comparative legal analysis. It seeks to identify prevailing regulatory models, pinpoint significant gaps in legal preparedness, and offer a normative basis for strengthening AI governance in telecommunications.

This paper makes four key contributions toward building resilient AI governance in critical infrastructure:

1. **Proposes a novel assessment instrument.** Develops the **AI Regulatory Readiness Index (ARRI)**, a structured, reproducible framework comprising seven indicators across three dimensions, scored on a 0–3 ordinal scale. As per our knowledge, ARRI is the first purely doctrinal legal readiness instrument specifically designed for AI governance in critical infrastructure, and it is generalizable beyond telecommunications.
2. **Establishes a transparent comparative methodology.** Introduces explicit instrument selection criteria, a five-tier legal hierarchy distinguishing binding law from soft-law instruments, and a structured search protocol for identifying relevant legal instruments across jurisdictions. Demonstrates that ARRI fills a gap not addressed by existing composite indices such as the GCI and GARI.
3. **Delivers a structured cross-jurisdictional analysis.** Applies ARRI across ten jurisdictions spanning five continents, producing a comparative scores matrix and a five-cluster typology of national regulatory approaches.
4. **Proposes a normative governance model.** Develops a minimum legal standards framework derived from ARRI scores, specifying the binding obligations that jurisdictions should enact to achieve baseline AI governance readiness in critical telecommunications infrastructure, with structured policy recommendations for regulators and legislators.

The remainder of this paper is structured as follows: Section 2 reviews the relevant literature and global readiness indices to identify research gaps. Section 3 details the methodology, describing the construction of the ARRI and its positioning vis-à-vis global frameworks, the scoring rubric, the instrument selection criteria and their hierarchy, and the criteria for jurisdictional and country selections. Section 4 analyses telecommunications, cybersecurity, data protection, and AI regulations across ten jurisdictions, providing the doctrinal basis for the readiness assessment. Section 5 presents the comparative results, featuring the ARRI score matrix and a typology of regulatory clusters. Section 6 evaluates these findings through the lens of regulatory theory and proposes a normative governance model for AI in critical infrastructure. Finally, Section 7 summarises the key contributions and potential impact of this work.

## 2. Literature review

The integration of Artificial Intelligence (AI) into critical infrastructure, such as telecommunications and energy, introduces autonomous decision-making that, while improving efficiency, creates novel risks including adversarial attacks, biases, and system failures with potential large-scale consequences [2]. Unpredictable machine learning models can fail in unforeseen situations, complicating safety and legal accountability [11], while algorithmic bias may perpetuate societal inequalities [12, 13]. AI's dual-use nature further complicates cybersecurity: it can be a powerful defensive tool, yet adversaries can weaponize it to launch automated attacks [14, 15, 16]. This threat is especially pronounced in 5G and future networks with expanded attack surfaces and large data volumes [17, 18], including AI-driven spam and social engineering in telecommunications [19]. These technical, ethical, and security challenges highlight gaps in existing frameworks [20].

While sector-specific AI governance scholarship has emerged in domains such as financial services [21] and healthcare [22], the telecommunications sector has received comparatively less attention as a distinct regulatory subject. A notable exception is recent work proposing a dedicated legal categorisation of "Telecommunication AI Incidents", distinguishing them from conventional cyber and data incidents to address failures specific to AI-enabled networks [8]. The present study builds on that doctrinal foundation by extending it into a comparative, cross-jurisdictional framework.

The proliferation of these complex risks has catalyzed a global consensus on the need for structured AI governance. Scholars argue that a formal normative framework is essential to mitigate risks affecting human rights, fairness, and safety, moving the discourse beyond abstract ethical principles toward practical implementation [23, 24]. AI governance is broadly understood as a system of rules, practices, and processes designed to ensure that an organization's use of AI technologies aligns with its strategies, objectives, values, and, critically, its legal and ethical obligations [25]. This has spurred development of various governance models across organizational, national, and international levels, each attempting to manage the risks inherent in the AI lifecycle [26].

To bridge the persistent gap between high-level principles and concrete enforcement, structured approaches are being proposed. Recent scholarship has outlined a five-layer AI governance framework, which systematically connects broad regulatory mandates with specific, verifiable implementation mechanisms like standards and certification [27]. This model positions formal laws and regulations as the foundational layer, providing the authoritative basis from which all other governance activities derive their legitimacy. This hierarchical view underscores the critical role of law in translating societal values and ethical considerations into enforceable requirements, a step deemed necessary as private self-regulation alone is considered insufficient to address the scale and complexity of AI-related challenges [23].

The imperative for AI governance raises a central policy dilemma between industry-led self-regulation and mandated legal frameworks [28]. While self-regulation is often valued for its flexibility and support for innovation, it is widely considered insufficient to ensure public accountability, enforceability, and the protection of fundamental rights [24, 29]. Consequently, scholarship increasingly supports a hybrid approach in which binding legal frameworks establish baseline safeguards, complemented by industry standards and internal governance mechanisms [30, 28, 31]. This approach reflects a broader shift toward embedding enforceable obligations within AI governance while retaining some flexibility for technological adaptation.

Closely related is the challenge of balancing innovation with ethical and legal safeguards [32]. While regulation is often viewed as constraining technological progress [33], this trade-off is increasingly seen as overstated [34]. Emerging scholarship argues that well-designed legal frameworks can promote trust and enable sustainable innovation [35]. This has led to calls for adaptive regulatory approaches that evolve alongside technological developments, enabling risk mitigation without imposing rigid constraints [36]. The focus thus shifts toward integrating legal and ethical considerations into innovation processes from the outset [32].

A significant deficiency in the current governance landscape is that many national legal frameworks predate the widespread integration of advanced AI, leaving critical regulatory gaps. Existing laws, particularly in the digital domain, were primarily designed to address harms like data breaches, unauthorized access, and network integrity, rather than the novel failure modes introduced by autonomous systems. The rapid pace of technological change has outstripped the adaptive capacity of these legal structures, creating a 'runaway world' scenario where governance lags behind risk [37]. Traditional laws are ill-equipped to handle harms stemming from algorithmic bias, lack of transparency, or unpredictable self-learning outcomes. This creates a 'responsibility gap', where the complex, opaque, and distributed nature of AI complicates the attribution of legal culpability [38]. As a result, governance activities such as bias testing, explainability, and safety culture remain largely recommended rather than legally mandated [39], leaving significant blind spots concerning emergent AI risks.

Scholars have identified structural reasons why legal frameworks lag behind technological risk. Regulatory path dependency suggests that existing legal architectures, built around traditional risks, such as data breaches and network intrusions, create institutional lock-in that makes AI-specific reform costly even when the need is recognised [40]. This is compounded by AI's horizontal nature, which cuts across sectors and creates overlapping authority among telecom regulators, data protection authorities, cybersecurity agencies, and emerging AI governance bodies [37]. Consequently, no single authority possesses both the mandate and technical capacity to govern AI-specific failures in critical infrastructure, a structural fragmentation that voluntary industry efforts cannot fully resolve [24].

In response to the identified regulatory vacuum, the private and non-profit sectors have initiated efforts to create a collective memory of AI failures [41]. The most prominent of these are voluntary AI incident repositories, such as the AI Incident Database (AIID), which aim to catalog real-world harms to prevent their recurrence and inform the development of more trustworthy systems [42]. These databases serve as an invaluable resource, providing empirical data on how AI systems can jeopardize safety and security in practice, from failures in autonomous vehicles to biases in facial recognition [43]. However, the voluntary nature of these initiatives reveals their inherent limitations. A significant body of research highlights that existing databases lack the standardized schemas and taxonomies required for consistent and structured data collection [44]. This absence of a unified protocol hinders systematic analysis, complicates cross-platform incident comparisons, and often requires expert inference to determine technical root causes rather than relying on structured data [45]. While these repositories are crucial for raising awareness and providing raw material for researchers [46], their shortcomings underscore the need for formalized, and potentially mandated, incident reporting frameworks to ensure that data is comprehensive, consistent, and actionable for regulatory oversight [47].

### 2.1. Existing Readiness Indices and the Legal Gap

Growing recognition of the need for structured AI governance has led to several cross-national readiness indices, each highlighting different aspects of national preparedness. The ITU Global Cybersecurity Index (GCI), now in its fifth iteration, evaluates national cybersecurity commitment across five pillars: legal, technical, organisational, capacity development, and cooperation [9]. Its legal pillar assesses the existence of cybercrime legislation, cybersecurity regulations, and incident response frameworks. However, the GCI does not assess AI-specific legal obligations, distinguish binding legislation from non-binding strategies, or provide sector-specific disaggregation.

The Oxford Insights Government AI Readiness Index (GARI) assesses government capacity to implement AI across three pillars (government, technology sector, and data and infrastructure), with a governance and ethics sub-pillar that captures AI strategy, regulation, and ethical frameworks [10]. While AI-focused, GARI combines legal with technical and institutional indicators, and does not separate enacted AI law from voluntary or advisory guidelines.

The UNESCO Readiness Assessment Methodology (RAM) provides a qualitative survey of government readiness to implement UNESCO's AI ethics recommendations across member states [48]. Although it includes a legal and regulatory environment dimension, it relies on government self-reporting and does not score binding instruments differently from soft-law documents.

Similarly, the OECD AI Policy Observatory maintains a comprehensive catalogue of national AI policies and regulations across its member and partner countries [49]. However, it is descriptive rather than scored, and does not assess sector-specific legal preparedness or the hierarchy of legal instruments. The Global AI Index by Tortoise Media benchmarks national AI capacity on investment, innovation, and implementation, with limited emphasis on legal preparedness [50].

Taken together, these indices reveal a consistent gap: none offers a reproducible, legally grounded assessment of AI governance readiness in critical infrastructure sectors, nor do they systematically distinguish binding laws from soft instruments. This study addresses that gap through the AI Regulatory Readiness Index (ARRI), a doctrinal instrument reproducible by researchers using primary legal instruments and applicable across critical infrastructure sectors, including telecommunications, healthcare, and transport.

The literature highlights a persistent gap between the emergent AI risks and existing governance capacities. National legal systems, largely predating widespread AI adoption, remain oriented toward traditional cybersecurity and data protection, leaving challenges such as algorithmic bias, autonomous system failures, and AI-specific incident reporting insufficiently addressed. Voluntary initiatives provide valuable empirical insights but also reveal the limitations of non-binding approaches. These findings underscore the need for structured, enforceable, and sector-sensitive mechanisms for effectively governing AI in critical infrastructure. This work addresses these gaps through ARRI.

## 3. Research Question and Methodology

This section outlines the research question and the multi-stage methodological approach used to develop and apply a novel instrument for assessing AI regulatory readiness in critical infrastructure. It details the construction of the *AI Regulatory Readiness Index (ARRI)*, the criteria for selecting legal instruments, and the rationale for focusing on the telecommunications sector and the ten jurisdictions included in this study.

### 3.1. Research Questions

This paper investigates the extent to which existing national legal and regulatory frameworks address the emerging risks posed by AI in critical infrastructure, with a particular focus on telecommunications. Specifically, the research seeks to answer the following questions:

*RQ1: To what extent do existing national legal frameworks address AI-related risks in critical digital infrastructure, particularly in telecommunications?*

*RQ2: How can a structured assessment instrument be designed to enable reproducible cross-jurisdictional comparison of legal preparedness for AI governance in critical infrastructure sectors?*

Answering these questions enables a systematic evaluation of whether current laws are equipped to handle AI-induced failures, vulnerabilities, or disruptions in telecom networks."

### 3.2. Overview of Methodological Approach

To address these research questions, this study employs a mixed-methods approach, combining qualitative legal analysis with a quantitative scoring framework. Existing relevant indices were reviewed to identify methodological gaps. A new index (ARRI) was then constructed, along with a clear and reproducible framework for instrument selection and legal hierarchy.

Ten jurisdictions representing diverse legal systems and geographies were selected for analysis. The study examines legal instruments across four domains: telecommunications, cybersecurity, data protection, and AI governance. Each instrument is qualitatively assessed to determine whether it explicitly or implicitly addresses AI-specific risks and incidents within telecommunications infrastructure. Numerical scores are assigned for each indicator based on this assessment.

Finally, these quantitative scores are aggregated to produce an overall ARRI score for each jurisdiction, enabling comparative analysis of regulatory readiness, identification of common patterns, and development of a typology of national governance models.

### 3.3. The AI Regulatory Readiness Index (ARRI)

The AI Regulatory Readiness Index (ARRI) is a structured, reproducible instrument for assessing the legal preparedness of national frameworks to govern AI in critical digital infrastructure. It is designed to operationalise the comparative legal analysis conducted in this study, transforming qualitative doctrinal assessments into ordinal scores that enable systematic cross-jurisdictional comparison.

The ARRI comprises seven indicators organised across three dimensions, each scored on a 0–3 ordinal scale. The aggregate raw scores are then normalised to a 0–100 index. The framework is inherently sector-portable: while this study applies it to telecommunications, the indicators and scoring logic are designed to evaluate any critical infrastructure domain, including energy, healthcare, and transport. The following subsections present the conceptual foundations of the index, the scoring rubric, and the positioning of the ARRI within existing global readiness frameworks.

#### 3.3.1. Indicators, Dimensions, and Design Rationale

The seven indicators that constitute the ARRI were derived directly from the regulatory gaps identified in the literature (Section 2) and operationalised through the legal domains examined in this study. Rather than adopting a generic governance checklist, each indicator targets a specific failure in existing legal frameworks that the literature identifies as inadequately addressed by traditional cybersecurity and data protection law. The indicators are organised into three dimensions that reflect distinct but complementary aspects of regulatory preparedness.

Dimension A captures the substantive content of legal obligations (the *what*); Dimension B assesses the operational safeguards and functional constraints placed on AI systems (the *how*); and Dimension C evaluates the institutional architecture required to oversee them (the *who*). Together, these three dimensions reflect the consensus in recent governance literature that effective AI regulation requires not only the right rules but also the structural capacity and the appropriate legal instruments to give those rules binding effect [24, 27].

- **Dimension A: Substantive AI-Specific Obligations (The "What")**: This dimension measures whether the legal framework explicitly recognises the unique failure modes of autonomous systems, distinct from traditional cyber/data breaches.
  - **Indicator A1: AI Incident Definition and Reporting**: Does the framework define AI-specific incidents (e.g., autonomous failures, model drift) distinct from cybersecurity incidents and mandate their reporting?
  - **Indicator A2: Mandatory Ex-Ante Evaluation**: Does the framework include provisions for ex-ante evaluation (e.g., impact assessments, certification, or mandatory testing) of AI systems before deployment or initial use?
  - **Indicator A3: Risk Classification and Management**: Does the framework mandate proactive, continuous risk management specifically tailored to AI systems (distinguishing high-risk critical infrastructure deployments from low-risk uses)?
- **Dimension B: Operational Safeguards for AI Systems (The "How")**: This dimension assesses whether the law imposes functional constraints on the design and operation of AI systems in critical infrastructure.
  - **Indicator B1: Algorithmic Transparency and Explainability**: Does the framework expect critical infrastructure operators to explain the logic of automated decision-making systems (e.g., network routing, spectrum allocation)?
  - **Indicator B2: Algorithmic Fairness, Reliability, and Auditing**: Does the framework establish requirements to test, audit, or mitigate risks such as bias, errors, or unsafe outcomes in AI systems deployed in critical operational contexts?
- **Dimension C: Governance Architecture and Coordination (The "Who")**: This dimension assesses the structural capacity of the state to oversee AI in complex, interconnected infrastructure, addressing the problem of "siloed" regulation.
  - **Indicator C1: Enforcement Authority and Accountability**: Is there a clearly designated regulatory authority (either a centralized AI body or an empowered sectoral regulator) with the statutory power to investigate AI failures and enforce compliance?
  - **Indicator C2: Cross-Domain Statutory Coordination**: Does the framework include formal coordination mechanisms between telecommunications regulators, cybersecurity agencies, and data protection authorities to prevent overlapping or contradictory AI oversight?

#### 3.3.2. Scoring Rubric

Each of the seven ARRI indicators is scored on a four-point ordinal scale from 0 to 3. The scale reflects the degree to which a jurisdiction's legal framework addresses the indicator through binding, enforceable obligations. A score of 3 represents the highest level of legal preparedness, where explicit and binding obligations exist in enacted law with designated enforcement authority. A score of 0 reflects complete legislative silence. The intermediate scores distinguish between binding obligations established through executive or ministerial administrative instruments (score 2) and non-binding soft-law instruments that address the indicator in principle but impose no legal obligation on regulated entities (score 1).

Equal weights are assigned to all seven indicators, as departing from equal weighting would require normative judgements about the relative importance of individual indicators that cannot be objectively grounded in the existing comparative legal literature. The composite ARRI score for each jurisdiction is calculated as follows:

$$\text{ARRI} = \left( \frac{\sum_{i=1}^{7} s_i}{21} \right) \times 100 \tag{1}$$

where $s_i$ denotes the score assigned to indicator $i$, each ranging from 0 to 3, and 21 is the maximum possible raw score. The resulting index is normalised to a scale of 0 to 100.

Table 1 presents the four scoring tiers, their definitions, and the corresponding instrument tiers. The scoring rubric explicitly links to the legal instrument hierarchy established in Section 3.4.2. Only instruments classified as Tier 1 (Enacted Statute) and Tier 2 (Delegated regulation) qualify for a maximum score of 3. Tier 3 instruments (Binding administrative instruments) result in a score of 2. Tier 4 soft-law instruments are capped at a score of 1 regardless of their substantive detail or scope, reflecting the absence of an enforceable legal obligation. Tier 5 draft or proposed instruments contribute no score, as unenacted bills impose no obligation on any regulated entity; their inclusion would risk conflating legislative aspiration with actual legal preparedness. Where a jurisdiction has instruments at multiple tiers addressing the same indicator, the score reflects the highest-tier binding instrument identified.

#### *3.3.3. Positioning ARRI within existing readiness framework*

The ARRI framework does not emerge in a vacuum. Several established global indices assess aspects of AI or cybersecurity governance at the national level, and an understanding of their scope and limitations is necessary to establish what ARRI contributes that existing instruments do not. Table 2 compares ARRI against five of the most relevant indices across four dimensions: the presence of a legal or regulatory component, the depth of doctrinal analysis, the ability to distinguish binding law from soft-law instruments, and adaptability to sector-specific critical infrastructure contexts.

The comparison reveals that no existing index simultaneously satisfies all four conditions that ARRI is designed to meet. The GCI comes closest in legal rigour but is confined to cybersecurity threats and does not assess AI-specific governance obligations. GARI and the Global AI Index measure governance capacity broadly but treat policy announcements and enacted law as equivalent signals of regulatory commitment. The OECD APO provides the most comprehensive instrument catalogue but does not score jurisdictions or distinguish the binding force of the instruments it records. UNESCO RAM engages most directly with legal frameworks but relies on self-reporting and prioritises ethical alignment over legal enforceability. ARRI is therefore designed to fill this specific methodological void. It combines doctrinal legal analysis (evaluating the text of laws), a nuanced scoring hierarchy (distinguishing binding statutes from policy aspirations), and a modular design (making it applicable across specific critical infrastructure sectors). It is proposed not as a replacement for these valuable macroeconomic and ethical indices, but as a complementary, legally-grounded instrument that operationalises the legal dimension of AI readiness in a form that existing composite indices acknowledge but do not currently provide.

### *3.4. Instrument Selection Criteria and Hierarchy*

Reproducible comparative legal analysis requires explicit rules governing which instruments are included, how they are ranked by legal force, and how they were identified. The following three components establish this foundation for the ARRI assessment.

#### *3.4.1. Inclusion criteria*

Five dimensions governed the selection of legal instruments for review across all ten jurisdictions. These criteria were applied uniformly, irrespective of jurisdiction, to ensure comparability and to prevent the inadvertent inflation of a jurisdiction's readiness score through the inclusion of non-binding or sector-irrelevant instruments.

1. **IC1: Legal nature (binding instruments preferred)**: Primary instruments must be enacted legislation, delegated/secondary regulations, or binding administrative rules with legal force. Non-binding instruments (strategies, guidelines, ethics codes, advisory circulars) are included only when no binding instrument exists on that topic, and are clearly flagged as soft law. Draft or proposed bills are included only to note legislative direction, never to support a readiness score.
2. **IC2: Domain relevance (four domains per jurisdiction)**: Instruments are selected from exactly four legal domains: (i) Telecommunications, (ii) Cybersecurity, (iii) Data protection, (iv) AI. This four-domain structure is constant across all jurisdictions, ensuring comparability.
3. **IC3: Temporal cut-off (instruments in force as of a stated date)**: Instruments formally enacted by 28 February 2026 are included for scoring purposes. Instruments that are formally enacted but subject to phased implementation or transition periods (e.g., the EU AI Act) are scored based on their final enacted text, as they already dictate immediate corporate compliance and investment trajectories. Draft bills/instruments are excluded from scoring.

Table 1: ARRI Scoring Rubric

| Score | Label | Definition | Qualifying instrument tiers |
|---|---|---|---|
| **3** | Established | Explicit, binding obligations exist in enacted primary legislation or binding subordinate regulation, backed by statutory enforcement authority. | Tiers 1–2 |
| **2** | Partial | Binding obligations exist through administrative instruments, operative ministerial orders, or regulatory decisions. | Tier 3 |
| **1** | Nascent | The indicator is addressed solely through non-binding instruments such as national strategies or ethical guidelines. | Tier 4 |
| **0** | Absent | The indicator is unaddressed by any active instrument. Any coverage is limited to unenacted draft legislation, imposing no current legal obligation. | Tier 5 or no instrument identified |

Table 2: Comparison of Established Global Indices with the Proposed AI Regulatory Readiness Index (ARRI)

| Index/ Methodology | Managing Organization | Core Focus | Legal/ Regulatory Component? | Doctrinal Depth (Distinguishes Binding vs. Soft Law)? | Sector-Specific Adaptability (e.g., Telecom, Energy)? |
|---|---|---|---|---|---|
| Government AI Readiness Index (GARI) | Oxford Insights | Public sector capacity, foundational data infrastructure, and technology skills. | Partial (Tracks national AI strategies and data laws). | No (Focuses on strategic intent rather than legal enforceability). | No (Macro-level focus on government operations). |
| Global AI Index | Tortoise Media | National AI capacity (Investment, Talent, R&D, and Operating Environment). | Partial (Legislation in Operating Environment pillar). | No (Quantitative tracking; treats policy announcements similarly to enacted legislation). | No (Macroeconomic focus). |
| Readiness Assessment Methodology (RAM) | UNESCO | Ethical and societal alignment with UNESCO's Recommendation on the Ethics of AI. | Partial (Legal environment dimension; relies on self-reporting, not independent analysis of enacted instruments). | No (Does not distinguish between binding legislation and non-binding strategies). | Partial (Can be adapted, but primary focus is broad societal and ethical impact, not sector-specific infrastructure governance). |
| AI Policy Observatory | OECD | Tracking and comparing national AI policies, strategies, regulations, and initiatives globally. | Yes (Comprehensive database of national strategies and guidelines). | Partial (Catalogues policies; does not evaluate bindingness). | No (General policy tracking). |
| Global Cybersecurity Index (GCI) | ITU | National commitment to cybersecurity (Legal, Technical, Organizational measures). | Yes (Evaluates cybercrime legislation and cybersecurity regulations). | Partial (Assesses existence of cybersecurity legislation but does not distinguish binding law from soft-law instruments). | No (General cybersecurity focus). |
| Proposed: AI Regulatory Readiness Index (ARRI) | Current Study | Evaluating the maturity, coherence, and enforceability of AI governance in critical infrastructure. | Yes (Focuses on the intersection of telecom, data, cyber, and AI law). | Yes (Explicitly scores jurisdictions based on the binding nature of instruments). | Yes (Designed as a modular framework applicable to specific critical infrastructures). |

4. **IC4: Primary sources (official government sources)**: Instruments are sourced from official government repositories (national gazette, parliament website, ministry publications) or authoritative translations thereof. Secondary commentary, news articles, or law firm analyses are used only to identify instruments, the instrument text itself is the unit of analysis.

5. **IC5: Scope (national/federal level primary)**: For federal systems (USA, Germany, Australia, Brazil, India), only federal/national-level instruments are included for the primary analysis. State-level instruments (e.g. California CCPA, German Länder data protection authorities) are excluded from base scoring unless they function as the de facto national regulatory standard in the absence of a federal equivalent (e.g., US data protection). This ensures comparability while reflecting actual compliance burdens.

*3.4.2. Legal instrument hierarchy*

Legal instruments vary substantially in their normative force, from enacted primary legislation to voluntary ethical guidelines. To ensure that ARRI scores reflect genuine legal obligation rather than mere policy aspiration, a five-tier hierarchy was established to classify each instrument before scoring.

1. **Tier 1: Enacted statute**: Primary legislation passed by the national legislature (e.g. Telecommunications Act, 2023 (India); EU Artificial Intelligence Act; Nigerian Communications Act 2003).
2. **Tier 2: Delegated regulation**: Secondary/subordinate legislation made under statutory authority (e.g. Telecommunications (Telecom Cyber Security) Rules, 2024 (India); China Network Data Security Regulations 2024).
3. **Tier 3: Binding administrative instrument**: Ministerial or regulatory instruments with binding legal force, typically sector-specific or technical in application, including orders, directions, regulatory decisions, and mandatory codes (e.g. Brazil ANATEL Resolution No. 767/2024; China Generative AI Measures 2023).
4. **Tier 4: Soft law/policy instrument**: Strategies, guidelines, voluntary codes, and ethical frameworks without binding legal force (e.g. Japan METI AI Governance Guidelines; Indonesia National AI Ethics Guidelines).
5. **Tier 5: Draft/proposed (pre-enactment instruments)**: Bills under parliamentary consideration and consultation drafts without current legal force (e.g. Brazil AI Bill (Senate-approved, Chamber pending); Japan AI Bill (Cabinet-approved, Diet pending)).

*3.4.3. Search and identification protocol*

Legal instruments were identified through systematic searches of official government sources, including national gazette repositories, ministry publications, legislative databases, and regulatory authority websites, across four domains: telecommunications, cybersecurity, data protection, and AI governance. Searches were conducted using domain-specific primary terms such as *Telecommunications Act*, *Cybersecurity Law*, *Data Protection Act*, and *Artificial Intelligence Act* or jurisdictional equivalents. Tier 1 enacted statutes were identified through parliamentary and legislative portals. Tier 2 subordinate instruments were located using terms such as *Rules*, *Regulations*, and *Implementing Regulation* combined with the relevant domain descriptor. Tier 3 administrative instruments were drawn from the official publications of sector regulators, including telecommunications commissions, cybersecurity agencies, and data protection boards. Tier 4 soft-law instruments, comprising national AI strategies, ethical guidelines, and advisory circulars, were sourced from competent ministry and government advisory body websites, and are included only where no binding instrument exists on a given indicator, with their non-binding status noted explicitly. Tier 5 draft bills were identified through parliamentary tracking portals and recorded solely to indicate legislative trajectory, contributing no score under ARRI. Secondary refinement terms applied across all tiers included *incident reporting*, *algorithmic accountability*, *explainability*, *risk classification*, and *critical infrastructure*. Where instruments were available only in a language other than English, authoritative government translations were used; secondary commentary was consulted solely to identify instruments, not to characterise their content. Industry standards issued by bodies such as ISO or ITU-T were excluded unless formally incorporated into binding national law.

### 3.5. Rationale for Focusing on Telecommunications

Telecommunications systems are chosen as the primary focus within critical digital infrastructure due to their foundational role in enabling connectivity, economic activity, and emergency services. As AI technologies become increasingly embedded in telecom operations, ranging from network management and resource allocation to fraud detection and customer service, they introduce new categories of risk that go beyond traditional cybersecurity or data protection concerns. These include unintended algorithmic behavior, systemic bias, lack of explainability, and emergent vulnerabilities due to AI's dynamic operation. Given the centrality of telecom networks and the rapid integration of AI into their functioning, analyzing their legal treatment offers a concrete and urgent perspective on legal readiness for AI risks in critical infrastructure.

### 3.6. Selection of Countries

Ten countries were selected for comparative analysis based on their global significance, geographic and legal diversity, technological maturity, and representation across five continents. These countries are: China, the United States, India, Japan, Indonesia, Brazil, Germany, United Kingdom, Nigeria, and Australia. Together, they represent a balanced mix of advanced economies, emerging markets, and varied legal traditions, offering valuable insights into global regulatory preparedness for AI risks in telecommunications. Table 3 presents their rankings in terms of mobile subscribers [51], GDP [52], and population [53], providing additional context for their inclusion.

Table 3: Selected countries by rankings and region

| # | Country | Mobile subscriber rank [51] | GDP rank [52] | Population rank [53] | Region | Brief Remark/Rationale |
|---|---|---|---|---|---|---|
| 1 | China | 1 | 2 | 2 | Asia | Global leader in 5G and AI adoption |
| 2 | USA | 3 | 1 | 3 | North America | Innovation leader; advanced telecom regulation |
| 3 | India | 2 | 5 | 1 | Asia | Rapid 5G expansion; new telecom legislation |
| 4 | Japan | 7 | 4 | 12 | Asia | Mature telecom ecosystem; high-tech economy |
| 6 | Indonesia | 4 | 16 | 4 | Asia | Largest Southeast Asian economy; regional case |
| 5 | Brazil | 8 | 9 | 7 | South America | Major Latin American emerging market |
| 7 | Germany | 18 | 3 | 19 | Europe | EU framework implementation; largest EU economy |
| 8 | United Kingdom | 21 | 6 | 22 | Europe | Post-Brexit regulation; mature telecom sector |
| 9 | Nigeria | 6 | 40 | 6 | Africa | Most populous African nation; growing mobile market |
| 10 | Australia | 49 | 13 | 55 | Oceania | Advanced regulation; key Asia-Pacific case |

## 4. National Legal Frameworks: A Cross-Jurisdictional Analysis

This section provides an overview of the existing legal and regulatory frameworks in ten countries, covering telecommunications, cybersecurity, data protection, and AI governance. It summarizes the key provisions in each domain to establish the current landscape of how AI-related risks in critical digital infrastructure, especially telecommunications, are addressed.

### *4.1. China*

China is the world's largest telecommunications market with a state-led regulatory model.

#### *4.1.1. Telecommunications*

The principal legal foundation for telecommunications in China is the *Telecommunications Regulations of the People's Republic of China (2016)*, originally issued in 2000 and last revised in 2016, which governs telecom licensing, operator obligations, interconnection standards, and network safety requirements [54]. Complementary instruments include the *Classified Catalogue of Telecommunications Services* (revised 2019) [55], the *Administrative Measures for the Licensing of Telecommunication Business* (2017) [56], and the *Regulations on the Administration of Foreign-Funded Telecommunications Enterprises* (2022) [57]. Together, these establish a comprehensive telecom regulatory framework covering licensing, service classification, market access, and operational requirements, but without explicit provisions for AI-driven network operations.

Incident reporting in the telecom sector is addressed through the *Emergency Response Plan for Data Security Incidents in the Industrial and Information Technology Sector (Trial)* (2024) [58], which mandates proactive risk prevention and requires reporting of major data security incidents to authorities within two hours. This framework was further consolidated by the *Administrative Measures for National Cybersecurity Incident Reporting*, issued by the Cyberspace Administration of China (CAC) in September 2025 and in force from 1 November 2025 [59]. These Measures unify previously dispersed reporting obligations across the CSL, DSL, and related regulations into a single framework with defined thresholds, timelines, and procedures applicable to all network operators, including telecom infrastructure operators. However, neither instrument defines or triggers reporting for AI-specific operational failures such as model drift or autonomous mismanagement.

#### *4.1.2. Cybersecurity, Data Protection, and AI*

China's cybersecurity framework is anchored in the *Cybersecurity Law (CSL)* (2017) [60], which protects critical information infrastructure, mandates network security measures, and governs cybersecurity incident management. Amendments adopted on 28 October 2025 and effective 1 January 2026 introduce Article 20, providing state support for AI research and development (algorithm design, training data and computing infrastructure, AI ethics, risk monitoring, and application promotion). They also align the CSL with the *Personal Information Protection Law (PIPL)* (2021) and the Civil

Code, requiring network operators to comply with these laws when processing personal information, and impose stricter penalties for cybersecurity and emergency response violations under amended Article 61.

The *Data Security Law* (2021) [61] introduces a data classification and grading system and mandates security reviews for data activities affecting national security. It is operationalised by the *Regulations on Network Data Security Management* (2024), in force from January 2025 [62], which impose compliance obligations on major network data processors. The *Personal Information Protection Law (PIPL)* (2021) [63] governs personal data processing, consent, data subject rights, and cross-border transfers. The *Measures for the Administration of Personal Information Protection Compliance Audits* (2025), effective 1 May 2025, introduce mandatory auditing for personal information handlers [64].

China has established an extensive AI regulatory framework. The *Administrative Provisions on Recommendation Algorithms* (2022) require algorithm-driven services to register (file) with the CAC, and in certain cases undergo security assessment, and are jointly administered by the CAC, the Ministry of Industry and Information Technology, the Ministry of Public Security, and the State Administration for Market Regulation [65]. The *Administrative Provisions on Deep Synthesis in Internet-based Information Services* (2023) regulate AI-generated content, including labelling and security requirements [66]. The *Interim Measures for the Administration of Generative Artificial Intelligence Services* (2023) mandate security assessments and filings for generative AI services prior to their launch [67] *(A2: 2)*. The *Measures for the Labelling of AI-Generated and Synthetic Content*, issued in March 2025 and effective 1 September 2025 [68], require explicit and implicit labelling of AI-generated content. Together, these instruments establish binding transparency and accountability obligations for AI systems, including those integrated into telecom platforms *(B1: 2, B2: 2)*.

At the strategic level, the *AI Plus* initiative (August 2025) [69] sets national targets for AI adoption across key sectors. The *AI Safety Governance Framework 2.0*, issued by TC260 in September 2025 [70], provides technical guidance for AI risk management and governance principles. China also issued a *Global AI Governance Action Plan* in July 2025 [71]. The Cyberspace Administration of China serves as the primary enforcement authority with powers to investigate, sanction, and ensure compliance *(A1: 1, A3: 2, C1: 2, C2: 2)*.

Across these instruments, incident reporting obligations apply to cybersecurity and data-related events, with stricter requirements for critical infrastructure and major data processors, but no framework defines AI-specific incident reporting triggers.

### *4.2. The United States of America*

The United States is a leading telecommunications market with a private sector-led AI ecosystem.

#### *4.2.1. Telecommunications*

U.S. telecommunications law rests on two foundational statutes. The *Communications Act of 1934* established the Federal Communications Commission (FCC) to regulate telephone, radio, television, cable, and satellite communications, governing frequency allocation, competition, public interest obligations, and national security provisions [72]. The *Telecommunications Act of 1996* modernised this framework to promote competition across converging communications markets [73]. Key amendments, such as the Communications Assistance for Law Enforcement Act (CALEA) and provisions under the USA PATRIOT Act, historically addressed lawful interception and national security. Regulatory safeguards for consumer privacy, nondiscrimination, and accessibility have also been embedded, balancing innovation and public interest. Neither statute addresses AI-specific risks in telecommunications operations, and neither has been amended for AI as of the study's cut-off date.

#### *4.2.2. Cybersecurity, Data Protection, and AI*

The United States has no comprehensive federal cybersecurity or data protection law, relying instead on a sector-specific and state-driven framework. The *Federal Trade Commission Act* empowers the Federal Trade Commission (FTC) to act against unfair or deceptive practices in cybersecurity and data protection [74]. The *Gramm-Leach-Bliley Act* mandates financial institutions to protect customer data, and its 2021 amendments to the Safeguards Rule strengthened cybersecurity obligations [75]. Sector-specific laws such as the *Health Insurance Portability and Accountability Act*, and Securities and Exchange Commission rules require data protection in healthcare and mandate disclosures of cybersecurity risks and incidents in financial markets, respectively [76, 77]. All 50 states have enacted data breach notification laws, with California's *Consumer Privacy Act (CCPA)* and *California Privacy Rights Act (CPRA)* representing the most comprehensive state-level framework [78].

The *Cybersecurity Information Sharing Act of 2015 (CISA 2015)* facilitates cyber threat information sharing between industry and government [79]. The Act lapsed on 1 October 2025, and its reauthorisation remained under legislative consideration as of 28 February 2026. The *Cyber Incident Reporting for Critical Infrastructure Act (CIRCIA)* of 2022 requires critical infrastructure operators, including telecommunications entities, to report significant cyber incidents to CISA within 72 hours [80]. The implementing rule, proposed in 2024, remained under review as of the cut-off date, with ongoing concerns regarding scope and alignment with existing reporting requirements. CIRCIA is exclusively a cybersecurity instrument and does not address AI-specific operational failures.

The United States has no federal AI law as of 28 February 2026. Federal AI governance operates primarily

through executive direction. President Trump's Executive Order 14179 (*Removing Barriers to American Leadership in Artificial Intelligence*), signed on 23 January 2025 [81], revoked the 2023 AI executive order and reoriented federal AI policy toward innovation and global leadership. The White House Office of Management and Budget issued a government-wide memorandum in April 2025 directing federal agencies to accelerate AI adoption, implement risk management practices for high-impact AI, and prioritise safe, secure, and resilient AI use [82] *(A2: 1, A3: 1)*. Executive Order 14365 (*Ensuring a National Policy Framework for Artificial Intelligence*), issued on 11 December 2025, directed the development of a minimally burdensome federal standard to preempt state AI laws inconsistent with federal innovation objectives, and instructed the FCC to initiate a proceeding within 90 days to consider adopting a federal AI reporting and disclosure standard for AI models [83] *(A1: 1)*. As of 28 February 2026, no such FCC proceeding had been finalised.

At the state level, AI legislative activity has accelerated, with all 50 states introducing AI-related bills in 2025 and numerous measures enacted across states [84]. Notable developments include New York City's Local Law 144 on bias audits, California's disclosure requirements for AI-generated content, and Colorado's algorithmic discrimination law (implementation delayed to June 2026) [84] *(B1: 0, B2: 1)*. The December 2025 Executive Order also established an *AI Litigation Task Force* [83] to challenge state laws deemed inconsistent with federal AI policy, contributing to regulatory uncertainty.

No federal or state law mandates AI-specific incident reporting across sectors, and no federal authority with AI-specific oversight of telecommunications has been established *(C1: 1, C2: 0)*. The FTC retains general enforcement authority over AI-related practices, while CIRCIA, if fully implemented, would cover cybersecurity incidents in telecommunications but not AI-specific operational failures.

### *4.3. India*

India is a large telecommunications market with a recently enacted Telecommunications Act and rapidly evolving digital and AI regulations.

#### *4.3.1. Telecommunications*

The primary legislation governing telecommunications in India is *The Telecommunications Act, 2023*, which received presidential assent in December 2023 and supersedes the Indian Telegraph Act, 1885 [85]. The Act governs the development, expansion, and operation of telecommunication services and networks, and the assignment of spectrum. It empowers the central government to issue rules to ensure the cybersecurity of telecommunication networks, designate any network as Critical Telecommunication Infrastructure, and prescribe corresponding security practices and standards. While the Act does not specify AI-related obligations, it provides an explicit enabling architecture for their future introduction, making it among the most forward-looking telecom laws globally.

The *Telecommunications (Telecom Cyber Security) Rules, 2024*, issued under the Act, mandate telecom entities to establish cybersecurity policies and respond rapidly to cyber incidents, with a six-hour notification requirement for security breaches and a detailed report within 24 hours [86]. The *Telecommunications (Critical Telecommunication Infrastructure) Rules, 2024* impose similar reporting obligations for security incidents in designated critical telecom infrastructure [87]. These rules establish a robust, binding incident reporting mechanism; however, their trigger is confined to cybersecurity breaches, and AI-specific failures such as model drift or autonomous network mismanagement fall outside their scope.

#### *4.3.2. Cybersecurity, Data Protection, and AI*

India's cybersecurity framework rests on *The Information Technology Act, 2000* (as amended in 2008) [88] supplemented by the Reasonable Security Practices and Procedures Rules and the CERT-In Rules, 2013 [89]. While they address cybercrime, unauthorized access, and cybersecurity incident management, they do not explicitly address AI-specific risks.

*The Digital Personal Data Protection Act, 2023 (DPDPA)* [90] establishes a comprehensive consent-based regime for personal data processing, incorporating purpose limitation, data minimization, and breach notification obligations. The *Digital Personal Data Protection Rules, 2025* [91] operationalise the Act through a phased compliance rollout and introduce Data Protection Impact Assessments and audits for Significant Data Fiduciaries *(A2: 2)*. Notably, the *Information Technology (Intermediary Guidelines and Digital Media Ethics Code) Amendment Rules, 2026*, notified on 10 February 2026 [92], introduce binding obligations for generative AI intermediaries, including content labelling and takedown requirements. While primarily platform-focused, they represent an early instance of AI-specific regulation under the IT Act framework.

India has no overarching AI legislation as of the study's cut-off date. The Telecom Regulatory Authority of India (TRAI) released non-binding recommendations in July 2023 proposing a risk-based regulatory framework for AI in telecommunications [93] *(A3: 1)*. The *India AI Governance Guidelines*, released in November 2025, provide a non-binding, principle-based framework and explicitly recommend establishing a national federated AI incident reporting mechanism and an India-specific AI risk classification framework [94] *(A1: 1)*. They propose three new institutions: an AI Governance Group (AIGG), a Technology and Policy Expert Committee (TPEC), and an AI Safety Institute (AISI) *(C1: 1, C2: 1)*. However, none of these institutions had been established by statute as of the cut-off date, and no binding AI rules had been notified *(B1: 1, B2: 1).* A Private Member's Bill on AI Ethics and Accountability, introduced in the Lok Sabha in December 2025, and a January 2026 white paper from the Office of the Principal Scientific Adviser advanced the techno-legal governance approach but introduced no new binding obligations [95, 96].

### *4.4. Japan*

Japan is an advanced telecommunications market with a cautious approach to AI regulation.

#### *4.4.1. Telecommunications*

The *Telecommunications Business Act (TBA)* forms the core of Japan's telecommunications regulatory framework [97]. It promotes fair competition, safeguards user interests, and ensures the secrecy of communications. The TBA also facilitates cybersecurity information sharing through mechanisms such as the ICT Information Sharing and Analysis Center Japan (ICT-ISAC Japan). Amendments enacted in June 2023 introduced external transmission rules governing the handling of user information on user devices, including cookies, and extended TBA jurisdiction to foreign telecommunications service providers. The TBA contains no AI-specific obligations or incident reporting requirements for failures arising from AI systems in telecommunications.

#### *4.4.2. Cybersecurity, Data Protection, and AI*

Japan's cybersecurity framework is anchored in the *Basic Act on Cybersecurity (BAC)* [98], which mandates national and local governments to enhance cybersecurity capabilities and encourages critical infrastructure operators to strengthen cyber defences through coordination mechanisms such as the Cybersecurity Council. A major legislative development occurred in May 2025 with the enactment of the *Active Cyber Defense (ACD) Act* [99], with phased implementation from 2026. The ACD Act introduces incident reporting obligations for critical infrastructure operators and enhances government authority to monitor foreign internet traffic traversing domestic infrastructure and respond to cyber threats *(C1: 1).* The reporting framework covers cyber incidents and the introduction of significant new IT systems but does not define or mandate reporting for AI-specific operational failures *(A1: 0).* Personal data protection is governed by the *Act on the Protection of Personal Information (APPI)* [100], substantially amended effective April 2023 to reinforce data breach notification requirements and expand data subject rights. Further revisions are under consideration, including potential implications for AI training data practices. On AI governance, Japan's regulatory approach underwent a significant transition in 2025 with the enactment of the *Act on Promotion of Research, Development, and Utilization of AI-Related Technologies* (AI Act) [101]. The Act promotes AI development while addressing associated risks, designates AI as a strategic priority, and establishes institutional mechanisms including the AI Strategy Headquarters and the formulation of an "Artificial Intelligence Basic Plan". It does not impose direct penalties or prescriptive obligations, instead adopting a best-efforts approach to compliance and relying on guidelines and existing sector-specific laws, including the APPI, competition law, and product safety law, for enforcement where necessary *(C2: 1).* Complementing the Act, the *AI Guidelines for Business* (version 1.1), jointly issued by METI and MIC in March 2025 [102], provide a non-binding framework for AI risk assessment, transparency, explainability, fairness, and incident handling. Structured across foundational values, cross-sector principles, and practical tools, the Guidelines are widely adopted by industry but do not impose legal obligations on telecommunications operators for AI-specific incident reporting or auditing *(A2: 1, A3: 1, B1: 1, B2: 1).*

### *4.5. Indonesia*

Indonesia is a major Southeast Asian telecommunications market with evolving digital regulation.

#### *4.5.1. Telecommunications*

Indonesia's telecommunications sector is governed by *Law No. 36 of 1999 on Telecommunications* [103], which established the framework for industry liberalisation, licensing, competition, and state oversight. *Government Regulation No. 52 of 2000* [104] elaborates licensing procedures and operational standards. The *Omnibus Law (Law No. 11 of 2020)* introduced cross-sectoral reforms, operationalised through *Government Regulation No. 46 of 2021 on Post, Telecommunication and Broadcasting* and *Minister Regulation No. 5 of 2021 on Telecommunications Operations* [105, 106], addressing spectrum sharing, infrastructure development, and pricing regulation. Article 22 of Law No. 36/1999 prohibits unauthorized acts on telecom networks, but does not cover AI-specific incidents. None of these instruments addresses AI integration or AI-specific risks within telecommunications.

#### *4.5.2. Cybersecurity, Data Protection, and AI*

Indonesia's cybersecurity and data protection framework is anchored in two primary statutes. The *Electronic Information and Transactions (EIT) Law* (Law No. 11 of 2008, as amended by Law No. 1 of 2024) [107] addresses cybercrimes including unauthorised access, defamation, and hacking, and governs electronic system operator obligations. The *Personal Data Protection (PDP) Law* (Law No. 27 of 2022) [108] establishes a comprehensive data protection regime, creating the Personal Data Protection Agency, mandating breach notification within 72 hours, and providing administrative and criminal sanctions for non-compliance, including rules on international data transfers. Implementing regulations under the PDP Law remain under development as of the cut-off date. *Government Regulation No. 71 of 2019 on*

*Electronic Systems and Transactions* [109] sets security and continuity obligations for electronic system operators, including telecommunications entities. *Presidential Regulation No. 47 of 2023* introduced the National Cybersecurity Strategy and Cyber Crisis Management Framework [110], complemented by *BSSN Regulations No. 1 and No. 2 of 2024* [111, 112]. BSSN Regulation No. 1 establishes procedures for cyber incident reporting and handling; it does not define or trigger reporting for AI-specific operational failures *(A1: 0).* Regulation No. 2 provides guidance on cyber crisis management *(C1: 1).*

The *Draft Law on Cyber Security and Resilience (RUU KKS)*, prioritised in the 2025 National Legislation Program and most recently revised in February 2025, aims to designate BSSN as the central cybersecurity authority, mandate incident reporting, and introduce sanctions for inadequate cybersecurity protection [113]. As of 28 February 2026, the bill remained in harmonisation at the Ministry of Law and had not been enacted *(Tier 5: no score).*

Indonesia has no umbrella AI law. The foundational policy document is the *National Strategy for Artificial Intelligence 2020–2045 (Stranas KA)* [114], which identifies five priority sectors and positions AI as central to the Golden Indonesia 2045 national development vision. The *Ministry of Communication and Informatics Circular Letter No. 9 of 2023 on AI Ethical Guidelines* [115] provides nine core values for AI developers, including transparency, accountability, and inclusiveness, but imposes no binding obligations *(A2: 0, B1: 1, B2: 1).*

In 2025, Indonesia's AI regulatory landscape advanced, though without enacting binding instruments. The Ministry of Communication and Digital (Komdigi) published the *National AI Roadmap White Paper* for public consultation in August 2025, establishing the policy architecture for a forthcoming Presidential Regulation on AI Ethics and Safety [116]. The draft regulation introduces a self-assessment mechanism for AI developers and ten ethical principles, including transparency, accountability, and personal data protection. Consultation concluded on 29 August 2025; the regulation was not yet promulgated *(Tier 5: no score).* The OJK issued *AI Governance Guidance for Indonesian Banking* in April 2025 [117], providing a lifecycle-based AI governance framework for the financial sector; it does not apply to telecommunications *(A3: 1).* Indonesia also participates in the ASEAN AI governance framework, including the *Expanded ASEAN Guide on AI Governance and Ethics* covering generative AI, issued in February 2025 [118], which provides non-binding regional principles. BSSN serves as the national cybersecurity authority; no AI-specific enforcement body exists, and no statutory mechanism coordinates AI, telecom, cybersecurity, and data protection oversight *(C2: 0).*

### *4.6. Brazil*

Brazil is a leading Latin American telecommunications market with developing AI governance.

#### *4.6.1. Telecommunications*

Brazil's telecommunications sector is governed by *Law No. 9,472 of 1997 (Lei Geral de Telecomunicações)* [119], also known as the *General Telecommunications Law.* It assigns federal responsibility for organising telecommunications services, managing spectrum and orbital resources, and promoting competition. It established the National Telecommunications Agency (ANATEL) as the sector's primary regulator, with powers to issue norms, conduct inspections, and apply sanctions. The General Telecommunications Law and subsequent ANATEL regulations do not explicitly address AI-related incidents or risks within telecom networks.

In April 2025, ANATEL approved the Regulatory Simplification project, which modernises existing rules and explicitly addresses the ethical and responsible implementation of AI in telecom services [120]. *Resolution No. 767 of 2024* updated ANATEL's cybersecurity framework, requiring telecom providers to assess cybersecurity vulnerabilities and report security incidents to ANATEL, aligned with mandatory notifications to the National Data Protection Authority (ANPD) [121]. The resolution expands coverage to submarine cable operators, mobile service providers, and wholesale network operators, and includes cybersecurity obligations related to cloud computing and data processing services. However, it does not define or mandate reporting for AI-specific operational failures *(A1: 0).*

#### *4.6.2. Cybersecurity, Data Protection, and AI*

Brazil's data protection framework is anchored in the *Lei Geral de Proteção de Dados Pessoais (LGPD)* (Law No. 13,709/2018, as amended by Law No. 13,853/2019) [122]. The LGPD governs personal data processing across sectors and establishes principles of lawful basis, purpose limitation, data minimisation, and data subject rights, including the right to request review of decisions based solely on automated processing. The National Data Protection Authority (ANPD), established in 2020, is the primary enforcement body.

*Resolution CD/ANPD No. 15/2024* [123] operationalises breach notification obligations under the LGPD, requiring notification within three business days for incidents involving sensitive or large-scale personal data. In 2025, the ANPD issued *Technical Note No. 12/2025* and the 2025–2026 Regulatory Agenda [124, 125]. It also conducted public consultations on artificial intelligence and automated decision-making under Article 20 of the LGPD. These developments signal ongoing efforts to regulate AI within the LGPD framework, though no binding AI-specific regulation has yet been adopted *(A2: 0, B1: 1, B2: 1).*

Brazil has no standalone cybersecurity law. Cybersecurity provisions are distributed across the *Internet Act (Marco Civil da Internet)*, the Consumer Protection Code, and the Criminal Code. The national cybersecurity strategy *E-Ciber* (2020) [126] provides strategic direction to strengthen cyber governance and promote crit-ical infrastructure protection, but does not impose binding obligations. CERT.br serves as the national Computer Emergency Response Team.

On AI governance, Brazil does not yet have a general law regulating the development and use of artificial intelligence. *Bill No. 2338/2023*, approved by the Federal Senate on 10 December 2024 and forwarded to the Chamber of Deputies in March 2025, proposes a rights-based, risk-tiered framework for AI governance [127]. The Bill adopts a risk-based approach, imposing stricter obligations on high-risk AI systems that may affect public safety or fundamental rights, including requirements for transparency, accountability, and human oversight. It also provides for the designation of a national supervisory authority and administrative penalties of up to BRL 50 million or 2% of annual revenue. The Chamber of Deputies established a special committee in April 2025 to examine the legislation, with deliberations continuing through late 2025 and a floor vote anticipated in early 2026. As of the study's cut-off date of 28 February 2026, the Bill had not been enacted *(Tier 5: no score)*. No binding AI-specific regulation exists as of the cut-off date, and no designated AI enforcement authority or cross-domain coordination mechanism has been established *(A3: 0, C1: 1, C2: 0)*.

### *4.7. Germany*

Germany reflects the national implementation of European Union's harmonised approach to digital and AI regulation.

#### *4.7.1. Telecommunications*

Germany's telecommunications sector is governed by the *Telecommunications Act (Telekommunikationsgesetz, TKG)*, comprehensively amended in 2021 to transpose the *European Electronic Communications Code (EECC)* [128]. The TKG regulates telecommunications services, network operations, and infrastructure management, with the Federal Network Agency (Bundesnetzagentur, BNetzA) as the central regulatory authority. It contains no explicit provisions addressing AI risks or mandating AI-specific incident reporting.

A major development during the study period is the *NIS2 Implementation and Cybersecurity Strengthening Act (NIS2UmsuCG)*, promulgated on 5 December 2025 (BGBl. I 2025 No. 301) and in force from 6 December 2025 [129]. The Act transposes the *EU NIS2 Directive* [130] into German law, substantially revising the BSI Act and amending sector-specific legislation including the TKG. Telecommunications providers are classified as essential or important entities, significantly expanding the scope of regulated entities to approximately 29,000. Covered entities must implement risk management measures, register with the Federal Office for Information Security (BSI), and comply with incident reporting obligations, detailed further in subsection 4.7.2. However, its incident reporting trigger is confined to cybersecurity incidents and does not extend to AI-specific operational failures.

At the European level, the *EU Artificial Intelligence Act* [131] introduces a risk-based framework for AI systems, including obligations on transparency, risk management, and incident reporting for high-risk AI systems, which will apply to telecommunications operators deploying such systems.

At the national level, BNetzA published a draft revised Security Catalogue for telecommunications networks and services in November 2025 for public consultation [132]. Once finalised, it will operationalise cybersecurity requirements for telecom operators, with a transition period before full implementation.

#### *4.7.2. Cybersecurity, Data Protection, and AI*

Germany's cybersecurity regime is centred on the *BSI Act (BSIG)*, under which the Federal Office for Information Security (BSI) serves as the lead authority. Expanded obligations were introduced by the *NIS2 Implementation and Cybersecurity Strengthening Act (NIS2UmsuCG)* [129], which revises the BSIG and amends the TKG, imposing enhanced risk management and incident reporting requirements on telecommunications and other critical sectors.

Germany's data protection framework is anchored in the *General Data Protection Regulation (GDPR)*, complemented by the national *Federal Data Protection Act (BDSG)* [133] and the *Telecommunications-Telemedia Data Protection Act (TTDSG)* [134], which governs communication confidentiality and metadata handling for telecommunications providers. The GDPR mandates Data Protection Impact Assessments for high-risk processing and breach notification to supervisory authorities. Enforcement is decentralised across the Federal Commissioner for Data Protection (BfDI) and Länder-level authorities.

Germany has no standalone national AI law but is directly subject to the *EU Artificial Intelligence Act (Regulation (EU) 2024/1689)* [131], which establishes a risk-based framework for AI systems. The Act applies in phases: prohibitions on unacceptable-risk AI practices from February 2025; obligations for general-purpose AI models from August 2025; and obligations for high-risk AI systems and full application by August 2026 *(A3: 3)*. High-risk AI systems must comply with requirements for data governance, risk management, technical documentation, transparency, human oversight, and cybersecurity. Telecommunications operators deploying AI systems for network optimisation, fraud detection, or customer service may fall within this category. They are subject to mandatory pre-deployment conformity assessment under Article 43, as well as registration and post-market monitoring obligations under other provisions of the Act *(A1: 2, A2: 3)*. The EU AI Act establishes national competent and market surveillance authorities. In Germany, responsibilities are shared across sectoral regulators, including the BSI and BNetzA, although no single dedicated AI supervisory authority had been designated as of the cut-off date *(B1: 3, B2: 2)*.

The integration of AI into telecommunications infrastructure creates overlapping obligations across cybersecurity, AI, and data protection regimes. Telecommunications operators must comply simultaneously with cybersecurity requirements under the BSIG and NIS2UmsuCG, AI obligations under the EU AI Act, and data protection duties under the GDPR, BDSG, and TTDSG. This multi-layered framework represents one of the most developed regulatory architectures in the study, though its full operationalisation depends on the finalisation of sector-specific instruments and the complete application of the EU AI Act in 2026 *(C1: 3, C2: 2)*.

### *4.8. The United Kingdom*

The United Kingdom adopts a principles-based, sector-led approach to AI governance.

#### *4.8.1. Telecommunications*

UK telecommunications is governed by the *Communications Act 2003* [135] and the *Wireless Telegraphy Act 2006* [136], with Ofcom as the national regulatory authority for licensing, spectrum management, and network resilience. The *EU's Electronic Communications Code (EECC)* was transposed into UK law before the end of the Brexit transition period, ensuring continuity in the telecom regulatory framework [137]. Secondary legislation enacted under the *EU Withdrawal Act 2018* corrected EU-specific references and enabled the domestic framework to remain functional and enforceable [138]. The *Telecommunications (Security) Act 2021* [139] strengthened network security obligations, empowering Ofcom to enforce security duties and manage risks from high-risk vendors. Its companion, the *Telecommunications Security Code of Practice* (December 2022) [140], provides detailed technical and contractual guidance, including incident response protocols, for major public telecom providers. Neither instrument includes AI-specific incident reporting obligations.

#### *4.8.2. Cybersecurity, Data Protection, and AI*

The UK's cybersecurity framework for essential services and telecom operators rests on the *Network and Information Systems (NIS) Regulations 2018* [141], which impose security and incident reporting duties on operators of essential services and digital service providers, but do not address AI-specific incidents. The *Cyber Security and Resilience (Network and Information Systems) Bill*, introduced to the House of Commons in November 2025 [142], proposes to expand the NIS 2018 regime by extending scope to managed service providers and data centres, introducing a 24-hour early warning requirement for significant incidents, and strengthening regulator enforcement powers including fines of up to £17 million or 4% of global turnover. As of the study cut-off date, the Bill was in Public Bill Committee and not enacted *(Tier 5, no score)*.

Data protection obligations arise under the *UK General Data Protection Regulation (UK GDPR)* and the *Data Protection Act 2018* [143, 144], mandating data breach notification and technical security measures for personal data. The *Data (Use and Access) Act* (2025) [145] updates UK data governance rules, affecting automated decision-making by relaxing some constraints on AI while introducing transparency requirements for certain automated processes *(B1: 1)*.

AI governance relies on a principles-based, sector-led approach articulated in the 2023 AI White Paper [146], which tasks existing regulators, Ofcom, ICO, FCA, and CMA, with applying cross-sectoral AI principles (safety, transparency, fairness, accountability, and contestability) within their existing powers *(A2: 1)*. The *AI Opportunities Action Plan* (January 2025) [147] reaffirmed this approach, prioritising AI investment and adoption. In February 2025, the AI Safety Institute was rebranded as the *AI Security Institute*, reflecting a clearer focus on serious AI risks with national security implications rather than broader issues such as bias [148]. The *Artificial Intelligence (Regulation) Bill [HL]* (March 2025) [149] proposes an AI Authority with binding powers, but remains a private member's bill without government backing. The *Code of Practice for the Cyber Security of AI* (January 2025) [150] provides voluntary guidance across twelve principles for managing AI cybersecurity risks, encouraging proactive risk assessment, monitoring, and incident management, but does not mandate formal reporting *(A1: 0, A3: 1, B2: 1)*. Cross-domain AI coordination is facilitated by the *Digital Regulation Cooperation Forum (DRCF)*, a voluntary advisory body comprising Ofcom, ICO, FCA, and CMA *(C1: 1, C2: 1)*.

### *4.9. Nigeria*

Nigeria is Africa's largest telecommunications market with an emerging regulatory framework.

#### *4.9.1. Telecommunications*

Nigeria's telecommunications sector is governed by the *Nigerian Communications Act (NCA), 2003*, which establishes the Nigerian Communications Commission (NCC) as the principal regulatory body with broad powers over licensing, spectrum allocation, infrastructure, interconnection, consumer protection, and quality of service [151]. Cybersecurity within the telecom sector is managed via institutional mechanisms, notably the NCC-CSIRT (Computer Security Incident Response Team), which provides vulnerability assessments, threat advisories, and incident response support to stakeholders [152]. Supporting regulatory instruments further establish technical and operational standards for service reliability, though none address AI-driven operations or AI-specific incident reporting.

The NCC has initiated reviews of telecommunications policy and subsidiary regulations, including stakeholder consultations on the National Telecommunications Policy. While these reviews may eventually incorporate emerging technologies such as AI and 5G, as of the cut-off date of this study, no formal amendments to the Nigerian Communications Act 2003 have been promulgated.

Strategic instruments include the *Nigerian National Broadband Plan 2020–2025* and the *National Digital Economy Policy and Strategy (2020–2030)* [153, 154], which recognise AI's role in modernising the telecom ecosystem but impose no binding obligations.

#### *4.9.2. Cybersecurity, Data Protection, and AI*

Nigeria's cybersecurity framework is anchored in the *Cybercrimes (Prohibition, Prevention, etc.) Act, 2015* [155], which criminalises hacking, identity theft, and unauthorized access, and includes critical infrastructure protections that may extend to AI-enabled systems interacting with networked resources. It also imposes obligations on service providers to retain user data and support law enforcement investigations. Cybersecurity coordination is guided by the *National Cybersecurity Policy and Strategy (NCPS, 2021)* [156], which underpins the Nigeria Computer Emergency Response Team (ngCERT) and provides guidance on sector-specific implementation plans, public awareness initiatives, and capacity building. Neither instrument defines or triggers reporting for AI-specific operational failures *(A1: 0)*.

On data protection, the *Nigeria Data Protection Act (NDPA, 2023)* [157] establishes the Nigeria Data Protection Commission (NDPC), codifies data subject rights, controller and processor obligations, and mandates breach notification to the NDPC and affected individuals. Section 37 prohibits fully automated decisions, including profiling producing legal or similar effects, without meaningful human oversight, providing a binding constraint on automated decision-making *(B1: 2)*. The NDPA builds on the *Nigeria Data Protection Regulation (NDPR, 2019)* [158], which set foundational principles for personal data handling. The NDPC's *General Application and Implementation Directive (GAID, 2025)* [159] provides enforcement rules for the NDPA *(C1: 1)*.

On AI governance, Nigeria has no enacted AI law as of the cut-off date. The *National AI Strategy*, revised in September 2025, outlines five pillars: modern infrastructure, a dynamic ecosystem, broad adoption, ethical deployment, and sound governance [160]. It emphasizes accountability, risk management, and ethical principles but remains non-binding *(A2: 0, A3: 1)*. A notable legislative development is the *National Digital Economy and E-Governance Bill, 2025* [161], which would designate NITDA as a central AI regulator with powers to classify AI risks, mandate transparency, and accredit AI auditors. However, as of the cut-off date of this study, the Bill had not been enacted *(Tier 5: no score, C2: 0)*.

### *4.10. Australia*

Australia follows a technology-neutral approach to AI within existing regulatory frameworks.

#### *4.10.1. Telecommunications*

Australia's telecommunications sector is governed by the *Telecommunications Act 1997* [162], amended in 2024. It mandates security obligations on network operators to protect networks from unauthorized access. The *Security of Critical Infrastructure Act 2018 (SOCI Act)*, also amended in 2024, designates over 220 telecom assets as systems of national significance [163]. The *Telecommunications Security and Risk Management Program (TSRMP) Rules*, effective April 2025, require carriers and carriage service providers with over 20,000 services to implement risk management programs and cybersecurity maturity frameworks [164]. The *Scams Prevention Framework Act 2025* requires telecom providers to adopt proactive anti-scam measures, thereby indirectly addressing AI-generated scam risks [165]. The *Telecommunications Amendment (Enhancing Consumer Safeguards) Bill 2025* [166] proposes to strengthen oversight through provider registration, enforceable industry codes, and enhanced penalties. However, these instruments do not explicitly address AI-specific risks in telecom operations.

#### *4.10.2. Cybersecurity, Data Protection, and AI*

The *Cyber Security Act 2024*, effective May 2025, introduces mandatory security standards for smart devices including telecom equipment [167]. It requires compliance statements, ransomware payment reporting, and establishes a Cyber Incident Review Board, indirectly addressing AI-enabled cyber threats in telecom. The *Security of Critical Infrastructure Act 2018* (amended 2024) [163] and Cyber Security Act impose binding incident reporting obligations on critical telecom operators, limited to cybersecurity threats *(A1: 0)*. The *Privacy Act 1988*, amended by the *Privacy and Other Legislation Amendment Act 2024*, strengthens data protection and introduces automated decision-making transparency requirements effective December 2026, applicable to AI systems processing personal data in telecom services [168] *(B1: 1)*. The Australian Privacy Principles (APPs) mandate data security and consent for AI-driven telecom applications but do not address AI-specific risks such as bias or accountability [169]. The *2023–2030 Australian Cyber Security Strategy* supports telecom cybersecurity through information sharing and risk management but lacks AI-specific provisions [170]. The *Consumer Data Right (CDR)* framework, expanding to telecom, promotes data portability without AI-specific rules [171].

Australia has no enacted AI law as of 28 February 2026. The government's AI governance trajectory shifted over the study period: after previously proposing mandatory AI guardrails for high-risk settings, the government released the *National AI Plan* on 2 December 2025, explicitly abandoning mandatory guardrails in favour of a technology-neutral approach relying on existing legal frameworks [172]. Rather than establishing new mandatory requirements, the plan commits to building on existing laws and regulatory regimes. The *Guidance for AI Adoption*, published by the National AI Centre in October 2025, condenses voluntary governance practices into essential practices for responsible AI deployment, including pre-deployment risk assessment [173] *(A2: 1, A3: 1)*. An *AI Safety Institute* was announced in November 2025 and committed funding for launch in early 2026, but had not been established or commenced operations as of the cut-off date [174] *(C1: 1)*. The *Digital Platform Regulators Forum (DP-REG)* provides non-statutory coordination among Australia's communications, consumer protection, privacy, and financial regulators [175] *(C2: 1)*.

## 5. ARRI Scores and Comparative Findings

This section presents the results of the application of the AI Regulatory Readiness Index (ARRI) to the ten jurisdictions under review. It moves beyond the narrative mapping in Section 4 to provide a structured, comparative assessment of how national legal frameworks are prepared to govern AI in the telecommunications sector.

### *5.1. ARRI Scores and Indicator Matrix*

Table 4 presents the ARRI scores for all ten jurisdictions across the seven indicators, together with each jurisdiction's raw score and normalised ARRI value. Scores reflect legal instruments enacted and in force as of 28 February 2026, in accordance with the instrument hierarchy and inclusion criteria established in Section 3.

The results span from 14 (Brazil) to 86 (Germany), revealing a highly uneven distribution of regulatory preparedness. The overall mean ARRI score (34) and median (26.5) indicate that most jurisdictions remain in the early to intermediate stages of AI regulatory development. Scores are concentrated in the lower range (19–38), with only one clear high-performing outlier, suggesting that comprehensive and mature AI governance frameworks remain limited.

Indicator-level analysis highlights systemic regulatory gaps. First, Indicator A1 (AI Incident Reporting) emerges as the weakest substantive obligation. Six jurisdictions score 0, and three score only 1, reflecting a near-universal reliance on legacy cyber-reporting thresholds that fail to capture AI-specific failure modes such as model drift. Second, Indicator C2 (Cross-Domain Coordination) represents a major governance constraint. Five jurisdictions score 0, while only two (China and Germany) achieve a score of 2, demonstrating that institutional mechanisms for coordinated oversight across telecommunications, cybersecurity, and data governance authorities remain underdeveloped.

At the dimension level, a consistent structural pattern emerges. Across jurisdictions, operational safeguards (Dimension B) are comparatively more developed than both substantive regulatory design (Dimension A) and governance coordination (Dimension C). Indicators relating to transparency and fairness are more widely implemented, whereas risk classification, ex-ante evaluation, and cross-domain coordination remain comparatively weak. This imbalance points to a broader tendency to prioritise technical or procedural safeguards over the development of integrated regulatory architectures.

The data also reveals differences in internal coherence across jurisdictions. Higher-scoring systems exhibit relatively balanced performance across all dimensions, whereas lower-scoring jurisdictions display fragmented profiles, with isolated indicators present but no sustained institutional depth. The absence of jurisdictions in the upper-middle range further reinforces the impression of a polarised landscape, in which regulatory systems tend to be either nascent or comparatively advanced rather than progressively intermediate.

### *5.2. Cross-Jurisdictional Patterns and Limits of Existing Indices*

The cross-jurisdictional patterns identified in the ARRI dataset provide a basis for comparison with existing global AI and digital readiness indices. In particular, the concentration of scores in the lower range, the imbalance across regulatory dimensions, and the prevalence of fragmented institutional profiles suggest that commonly used composite indices may overstate the maturity and coherence of national AI governance systems.

The ARRI scores diverge systematically from the rankings produced by established composite readiness indices, revealing a fundamental measurement gap in the existing global AI governance assessment landscape. The United States ranks among the highest jurisdictions in the Oxford Insights Government AI Readiness Index (GARI), yet scores 24 under ARRI. Similarly, Indonesia achieves the highest possible status (Tier 1: Role-modelling) in the ITU Global Cybersecurity Index (GCI) 2024 with a score of 100 out of 100, yet records an ARRI score of 19, placing it in the lowest band.

These divergences are not anomalies. **They reflect a core analytical finding of this study: cybersecurity readiness and AI regulatory readiness are legally and functionally distinct conditions that existing composite indices structurally conflate.** High GCI scores capture mature cybersecurity legislation, incident response capabilities, and national strategies, but do not address AI-specific risks such as model drift, algorithmic bias, or autonomous system failures. Similarly, indices such as GARI measure strategic intent and policy articulation without distinguishing between binding regulatory frameworks and soft-law instruments. ARRI operationalises precisely this distinction, which explains the divergence in rankings.

Table 4: The AI Regulatory Readiness Index (ARRI) score matrix across ten jurisdictions

| Jurisdiction | A1 | A2 | A3 | B1 | B2 | C1 | C2 | Raw /21 | ARRI /100 |
|---|---|---|---|---|---|---|---|---|---|
| China | 1 | 2 | 1 | 2 | 2 | 2 | 2 | 12 | 57 |
| USA | 1 | 1 | 1 | 0 | 1 | 1 | 0 | 5 | 24 |
| India | 1 | 2 | 1 | 1 | 1 | 1 | 1 | 8 | 38 |
| Japan | 0 | 1 | 1 | 1 | 1 | 1 | 0 | 6 | 29 |
| Indonesia | 0 | 0 | 1 | 1 | 1 | 1 | 0 | 4 | 19 |
| Brazil | 0 | 0 | 0 | 1 | 1 | 1 | 0 | 3 | 14 |
| Germany | 2 | 3 | 3 | 3 | 2 | 3 | 2 | 18 | 86 |
| UK | 0 | 1 | 1 | 1 | 1 | 1 | 1 | 6 | 29 |
| Nigeria | 0 | 0 | 1 | 2 | 0 | 1 | 0 | 4 | 19 |
| Australia | 0 | 1 | 1 | 1 | 0 | 1 | 1 | 5 | 24 |
| *Mean* | | | | | | | | 7 | 34 |
| *Median* | | | | | | | | 5.5 | 26.5 |

*Scores:* 0 = Absent; 1 = Nascent; 2 = Partial; 3 = Established. *Indicators — Dimension A (Substantive):* A1 = AI incident reporting; A2 = Ex-ante evaluation; A3 = Risk classification; *Dimension B (Operational safeguards):* B1 = Transparency and explainability; B2 = Fairness, reliability, and auditing. *Dimension C (Governance):* C1 = Enforcement authority; C2 = Cross-domain coordination. ARRI = (Raw/21) × 100, rounded to nearest integer.

The structural explanation for this pattern lies in **regulatory path dependency**. Legal frameworks built around pre-existing risk categories, such as cybersecurity incidents and data breaches, create institutional lock-in that renders AI-specific reform administratively and politically costly, even where the need is recognised. The horizontal nature of AI as a general-purpose technology compounds this challenge by generating overlapping authority across telecommunications regulators, data protection authorities, cybersecurity agencies, and emerging AI governance bodies. As a result, in most jurisdictions studied, no single authority possesses both the statutory mandate and the technical capacity to govern AI-specific failures in telecommunications. This structural limitation cannot be fully addressed through voluntary industry initiatives or soft-law frameworks [24, 37].

### 5.3. *Typology of Regulatory Clusters*

Building on the observed variation in ARRI scores and indicator-level patterns, the ten jurisdictions are grouped into five distinct regulatory archetypes. This typology reflects not only differences in overall readiness but also the underlying structural approaches to AI governance in critical infrastructure.

First, **Comprehensive Regulators** (e.g., Germany) demonstrate high and balanced performance across all dimensions. Well-developed substantive rules, robust operational safeguards, and strong governance coordination characterize these systems. They reflect institutionalised, system-wide approaches to AI regulation, where regulatory design, implementation, and enforcement are closely aligned.

Second, **Coordinated Transition Regulators** (e.g., China) exhibit relatively strong and consistent performance across indicators, but primarily at an intermediate level. These systems demonstrate broad regulatory coverage and coordination, often driven by centralised or state-led approaches, yet lack the granular, sector-specific depth associated with fully mature frameworks.

Third, **Emerging Structured Regulators** (e.g., India, UK, Japan) include jurisdictions that have established foundational regulatory frameworks across multiple dimensions but remain at an early stage of institutionalisation. Their profiles are characterized by consistent but low-to-moderate scores, reflecting the presence of guiding principles and recent statutory developments, although they lack fully developed enforcement or reporting mechanisms.

Fourth, **Operationally Skewed or Sectoral Regulators** (e.g., United States, Australia) display uneven development, with relatively stronger performance in operational safeguards (Dimension B) but weaker integration in substantive regulation (Dimension A) and cross-sectoral governance (Dimension C). These systems often rely on decentralised, sector-specific, or industry-led approaches, resulting in fragmented but functionally active regulatory landscapes that prioritise transparency over binding obligations.

Finally, **Nascent or Fragmented Regulators** (e.g., Indonesia, Nigeria, Brazil) are characterised by low scores across most indicators and limited institutional coherence. Regulatory efforts in these jurisdictions tend to be ad hoc or reactive, with minimal coordination and limited statutory capacity for comprehensive AI oversight in critical infrastructure.

This fivefold typology highlights that differences in AI regulatory readiness are not merely a matter of degree but reflect qualitatively distinct governance models. In particular, it underscores that coordination (C2) and incident reporting (A1) emerge as key differentiating factors between more and less mature regulatory systems.

## 6. Discussion

This section evaluates the ARRI findings by identifying systemic governance gaps and archetype-specific challenges. On this basis, it proposes a normative model for baseline readiness and offers actionable policy recommendations to address regulatory fragmentation and strengthen AI oversight in critical telecommunications infrastructure.

### *6.1. Governance Gaps and Uncovered AI Risks in Telecommunications*

The ARRI findings reveal persistent governance gaps that translate into unaddressed systemic risks within telecommunications infrastructure. These gaps represent substantive omissions that limit the ability of legal systems to detect, assess, and respond to AI-related failures in critical network operations.

The most consequential gap concerns the absence of binding AI-specific incident reporting obligations. Most regulatory frameworks continue to rely on legacy cybersecurity and service disruption thresholds, which are not designed to capture AI-induced failures such as model drift, optimisation errors, or unintended system interactions. Consequently, failures in AI-driven network management may remain invisible to regulators unless they escalate into conventional outages or security breaches. This creates a systemic blind spot in oversight, preventing the accumulation of institutional knowledge and the feedback mechanisms necessary for adaptive governance and risk mitigation [45, 176].

A second gap lies in the limited application of risk classification frameworks. In most jurisdictions studied, AI applications in core network functions are not formally designated as high-risk, meaning they do not trigger ex ante obligations such as mandatory impact assessments or human oversight. This omission is significant given the systemic role of telecommunications, where failures in AI-enabled routing or spectrum management can propagate rapidly across interconnected infrastructure. The absence of risk-tiered obligations reduces the capacity of legal frameworks to anticipate and mitigate failures prior to deployment.

A further structural constraint arises from weak cross-domain coordination. Telecommunications, cybersecurity, data protection, and emerging AI governance frameworks often operate in parallel, with limited statutory mechanisms for integrated oversight. In practice, a single AI-related failure may fall within multiple regulatory domains without any authority having a comprehensive mandate to assess its impact. This institutional separation reflects the challenge of governing AI as a general-purpose technology within sector-specific legal architectures that were not designed for its cross-cutting effects.

The implications of these gaps vary across the regulatory archetypes identified in this study. In jurisdictions with nascent or fragmented systems, the primary risk is the absence of enforceable mechanisms to govern AI deployment in critical infrastructure. In more operationally developed but decentralised systems, the risk is one of partial coverage, where transparency and auditing measures are present but are not supported by substantive obligations that shape system design and deployment. In both cases, the result is a form of regulatory insufficiency, in which existing frameworks are unable to fully constrain the risks introduced by AI in telecommunications.

### *6.2. A Normative Model for Critical Infrastructure AI*

The ARRI analysis provides an empirical basis for articulating minimum standards of AI regulatory preparedness in critical telecommunications infrastructure. This subsection proposes a normative model derived from the study's scoring results and the five-archetype typology.

The model posits that a jurisdiction achieves **baseline AI governance readiness** when it scores 2 or above on all seven ARRI indicators, yielding a minimum raw score of 14/21 and a normalised value of approximately 67. Under this threshold, only Germany currently meets baseline readiness. This threshold is achievable without a purpose-built horizontal AI Act; it requires the strategic activation of existing enabling architectures, the amendment of foundational telecom laws, and the statutory empowerment of regulatory institutions.

The reference architecture for baseline readiness comprises three interlocking layers:

1. **Substantive layer (Dimension A):** binding obligations for AI incident reporting, ex-ante evaluation, and risk classification.
2. **Operational layer (Dimension B):** mechanisms ensuring transparency, reliability assurance, and auditing requirements.
3. **Governance layer (Dimension C):** statutory designation of enforcement authority with AI-specific powers and formal cross-domain coordination mechanisms.

These layers are mutually reinforcing: substantive obligations without enforcement are unenforceable; enforcement authority without coordination produces siloed oversight; and coordination without substantive obligations lacks meaningful scope.

Adaptive governance mechanisms are critical to the model. Given the rapid evolution of AI, regulatory frameworks must establish feedback loops from incident reporting, periodic reassessment of risk classifications, and iterative updates to operational safeguards in response to emerging failure modes such as model drift.

The normative priorities differ by regulatory archetype:

- **Nascent or Fragmented Regulators:** establish the governance layer by designating a national AI coordination authority and enacting foundational reporting obligations.

- **Emerging Structured Regulators:** activate existing enabling architecture through binding subordinate regulations or sector codes.

- **Operationally Skewed Regulators:** close the gap between operational safeguards and substantive mandates by enacting risk classification frameworks.

- **Coordinated Transition Regulators:** extend existing algorithmic instruments from content governance to oversight of critical network functions.

- **Comprehensive Regulators:** maintain balanced readiness across all layers and continuously adapt governance mechanisms in response to emerging AI risks.

### *6.3. Policy Recommendations*

Five recommendations address the specific indicator gaps identified by ARRI and are structured to be actionable within existing legislative architectures:

1. **Enact binding AI incident reporting obligations in telecommunications law (A1).** Jurisdictions relying on legacy cybersecurity reporting thresholds should amend their telecom security rules to define AI operational failures as a distinct reportable category, covering events such as model drift, autonomous mismanagement, and adversarial manipulation of AI-driven network functions [8]. Subordinate legislation under existing enabling powers can implement these requirements, enabling rapid activation without waiting for comprehensive AI Acts. A harmonised taxonomy of Telecommunication AI Incidents, developed through ITU standardisation processes, would support cross-border consistency in reporting and data collection.
2. **Mandate ex-ante risk evaluation for AI systems in critical telecom operations (A2, A3).** Jurisdictions currently lacking enacted risk classification should adopt tiered frameworks designating AI systems used in core network management, emergency services routing, and critical infrastructure control as high-risk. This designation should trigger pre-deployment conformity assessment and human oversight obligations without requiring comprehensive AI legislation. The EU AI Act's risk classification methodology provides a replicable model adaptable to national legal systems.
3. **Introduce binding transparency and fairness assurance obligations for AI in telecom (B1, B2).** Jurisdictions relying solely on data protection law for automated decision-making transparency should extend these obligations explicitly to AI systems in telecom network operations, not only to personal data processing. Binding algorithmic auditing requirements should be enacted for AI systems making consequential decisions in spectrum allocation, fraud detection, and customer service, with audit results reported periodically to the designated telecommunications regulator.
4. **Designate or empower a statutory authority with AI-specific oversight of telecommunications (C1).** Existing telecom regulators should be empowered through statutory amendment to investigate AI operational failures, impose sanctions for non-compliance with AI governance obligations, and coordinate with data protection and cybersecurity authorities on AI-related matters.
5. **Mandate statutory cross-domain coordination for AI oversight in telecommunications (C2).** Voluntary coordination forums should be converted into statutory bodies with defined mandates, membership, and decision-making procedures. These bodies should develop joint guidelines for AI incident classification, coordinate enforcement responses to AI failures that span regulatory domains, and report annually to the legislature on the adequacy of the national AI governance framework for critical infrastructure.

Across all five recommendations, the appropriate balance between safety obligations and innovation enablement must be maintained. Prescriptive obligations imposed without technical guidance risk compliance-without-comprehension among telecom operators. Each binding obligation should therefore be paired with a corresponding guidance instrument, such as sector-specific codes of practice, technical standards, or regulatory sandbox mechanisms, supporting implementation without restricting AI-enabled efficiency gains. Where relevant, recommendations should also align with international standards and reporting practices to promote cross-border consistency and resilience in critical infrastructure networks.

### *6.4. Limitations and Future Work*

This study has several limitations that should be noted. First, the analysis focuses on ten jurisdictions selected for diversity in legal traditions and geography, but they do not represent the global landscape. Important developments in jurisdictions such as Singapore, South Korea, Canada, and smaller EU member states are not captured. Extending ARRI to a broader sample would test the generalisability of the five-archetype typology. Second, the ARRI assessment reflects a single temporal snapshot as of 28 February 2026. Rapidly evolving legislation, such as Brazil's pending AI Bill or Japan's forthcoming AI Basic Plan, may materially alter regulatory readiness in the near future. Third, ARRI operationalises the distinction between binding and non-binding instruments as a binary. In practice, the compliance effect of a widely adopted voluntary standard may exceed that of a poorly enforced binding rule. ARRI thus captures legal form rather than practical enforcement outcomes, and should ideally be complemented by studies of actual regulatory behaviour. Fourth, instruments available only in non-English languages were interpreted via official translations, which may introduce minor artefacts affecting scoring.

Future research should address four priorities. (i) ARRI could be applied to other critical infrastructure sectors, such as energy, healthcare, and transport, to assess whether the governance gaps observed in telecommunications are systemic across critical infrastructure or sector-specific. The framework's modular design supports this extension without structural modification. (ii) Repeated application of ARRI to the same ten jurisdictions at two-year intervals would allow tracking of regulatory maturity over time and evaluation of whether jurisdictions are converging towards baseline readiness. (iii) The development and standardisation of AI incident repositories in critical infrastructure would provide empirical data to refine scoring criteria and benchmark the effectiveness of mandatory reporting frameworks. (iv) Engagement with international standardisation bodies, including the ITU and the Global Partnership on Artificial Intelligence, could promote harmonised minimum standards for AI governance, reducing cross-border regulatory fragmentation and supporting resilient multinational operations.

## 7. Conclusion

This study examined the legal and regulatory frameworks governing Artificial Intelligence (AI) in critical telecommunications infrastructure across ten jurisdictions spanning five continents. To address the methodological gap in existing literature, it developed and applied the **AI Regulatory Readiness Index (ARRI)**, a structured, reproducible instrument designed for the doctrinal assessment and cross-jurisdictional comparison of AI governance readiness.

Four principal conclusions emerge from this comparative analysis. First, the global regulatory landscape for AI in telecommunications is characterised by pervasive regulatory insufficiency. The mean ARRI score of 34 (out of 100) and the median of 26.5 indicate that the central tendency of legal preparedness remains in the lower range of regulatory maturity, reflecting a widespread condition in which jurisdictions may possess operational guidelines but lack comprehensive, binding AI governance frameworks. Second, AI incident reporting (Indicator A1) and proactive risk classification (Indicator A3) represent the most acute universal gaps. Six jurisdictions score 0 on A1, confirming the absence of binding legal definitions of AI-specific incidents in telecommunications law across most jurisdictions studied. Consequently, the regulatory mechanisms most essential to proactive AI oversight remain largely underdeveloped. Third, cross-domain coordination (Indicator C2) is a defining structural weakness. The institutional mechanisms required for integrated oversight across telecommunications, cybersecurity, and data governance authorities are absent in statutory form across half the jurisdictions studied, highlighting a fundamental mismatch between AI's horizontal nature and sector-specific regulatory design. Fourth, ARRI scores diverge systematically from established composite readiness indices. This divergence confirms the study's core finding: cybersecurity readiness and AI regulatory readiness are legally and analytically distinct conditions that existing composite indices structurally conflate.

To systematically interpret these divergences, the study classified the jurisdictions into a **five-archetype typology** of regulatory maturity, ranging from Comprehensive Regulators to Nascent or Fragmented Regulators. This typology demonstrates that differences in AI governance are not merely quantitative but reflect distinct structural approaches to coordination, operational safeguards, and substantive obligations.

The normative contribution of this study is therefore twofold. First, ARRI provides a reproducible, doctrinal assessment instrument that enables cumulative comparative legal scholarship on AI governance across critical infrastructure sectors. Second, the **normative governance model** proposed in Section 6, anchored at an ARRI score of 67 as the baseline for readiness, offers a concrete reform target achievable within existing legislative architectures. As AI becomes the operational backbone of critical digital infrastructure, legal frameworks must move from a reactive to an anticipatory posture. The model and recommendations advanced in this study provide both the analytical framework and the normative direction required to address the prevailing governance gap and support the resilient deployment of AI in critical telecommunications networks.